\documentclass[aps,prl,twocolumn,floatfix]{revtex4}
\usepackage{graphicx}

\begin{document}

\title{Fluctuations of a homeotropically aligned nematic liquid crystal in the presence of an applied voltage}
\author{Sheng-Qi Zhou and Guenter Ahlers}
\address{Department of Physics and iQUEST, University of California, Santa Barbara, California 93106}
\date{\today}
\begin{abstract}
We determined the refractive-index structure-factor $S_n(\bf k)$ from shadowgraphs of fluctuations in a layer of a homeotropically aligned nematic liquid crystal with negative dielectric anisotropy in the presence of an ac voltage of amplitude $V_0$ applied orthogonal to the layer. $S_n(\bf k)$ had rotational symmetry. Its integral $P(V_0)$ and amplitude $B(V_0)$ increased smoothly through the Fr\'eedericksz transition at $V_0 = V_F$. Its inverse width $\xi(V_0)$ and its relaxation rate $\Gamma_0(V_0)$ had cusps but remained finite at $V_F$. The results are inconsistent with the critical mode at a second-order phase transition.

\end{abstract}

\maketitle

A nematic liquid crystal (NLC) consists of elongated molecules that, because of their shape,  tend to align locally relative to each other. \cite{Bl83} The alignment direction is called the director $\hat n$. When a NLC is confined between parallel glass plated  with a small spacing $d$ between them, $\hat n$ is influenced by the interaction of the molecules with the glass surfaces. It is possible to prepare the surfaces in such a fashion as to cause more or less uniform director alignment throughout a thin sample. When $\hat n$ is orthogonal (parallel) to the surfaces,  the alignment is called homeotropic (planar). We note that planar alignment determines a preferred direction in the plane of the sample and thus breaks the rotational invariance characteristic of an isotropic fluid. However, in the homeotropic case, which is of interest here, this rotational symmetry is preserved.

Most physical properties of a NLC are anisotropic. We applied a voltage $V = \sqrt{2}V_0\cos(\omega t)$ between transparent indium-tin oxide (ITO) electrodes on the inner surfaces of the confining plates, and thus are interested in the anisotropy of the dielectric constant $\epsilon_a = \epsilon_{\parallel} - \epsilon_{\perp}$ where  $\epsilon_{\parallel}$ and $\epsilon_{\perp}$ are the dielectric constants parallel and perpendicular to the director. When $\epsilon_a < 0$, the homeotropic state becomes unstable when $V_0$ exceeds a threshold value $V_F$. 
The transition is known as the Fr\'eedericksz transition (FT). \cite{FZ33}  Above $V_F$ the director, in the ideal case, remains homeotropically anchored at the surfaces, but away from the confining plates it acquires a component $n_x$ in the plane (we arbitrarily choose the 
$x$-axis in the direction of this deformation). 

A stability analysis at the mean-field level of the FT was given by Hertrich {\it et al.}\cite{HDPK92} They found that the transition is of second order and that it occurs first for wave number $k = 0$. 
Here we are concerned with the time-dependent fluctuations of the director field \cite{WRRZKB91,ELVJ89, GR94} that are induced by thermal noise and not seen in global or time-averaged measurements \cite{MPB86}.  A spatial variation of the component of $\hat n$ in the plane of the sample is associated with a variation of the vertical average of the refractive index $n$ and thus  can be seen by the shadowgraph method. \cite{RHWR89,BBMTHCA96} For the purpose at hand the method has the advantage that it sees only the fluctuations and not the much larger spatially uniform change of the refractive index above the transition that influences the total transmitted intensity when the sample is viewed between crossed polarizers.  \cite{MPB86} This is so because its sensitivity at small $k$ is proportional to $k^2$ and thus vanishes for $k = 0$. From shadowgraph images we derived the structure factor $S_n({\bf k},V_0)$ of the refractive-index field using the optical transfer function derived from physical optics.\cite{TC02}

We looked for, but did not find, any hysteresis at $V_F$. For fluctuations near the critical point  of an equilibrium system \cite{St71} or a supercritical bifurcation of a non-equilibrium system \cite {SH77,RRTSHAB91,WAC95,OA03,OOSA04} the fluctuation power $P$ of the critical mode should pass through a sharp maximum as  a control parameter $\epsilon$, in our case equal to $V_0^2/V_F^2-1$, increases from negative to positive values. Such a maximum was observed for instance near the onset of RBC.\cite{OA03} We found that $P$ varied smoothly and {\it monotonically} through $\epsilon = 0$, continuing to {\it increase} with $\epsilon$ above the transition. The correlation length $\xi(V_0)$, equal to the inverse width at half height of the azimuthal average $S_n(k)$ of $S_n({\bf k})$, is expected to diverge at $\epsilon = 0$. We found that it had a sharp but finite cusp, indicating a well defined transition with ``rounding" confined to $|\epsilon| \alt 0.01$. The relaxation rate $\Gamma_0(V_0)$ is expected to vanish at $V_F$; it was found to have a sharp down-pointing cusp at $V_F$  but remained non-zero. The height $B$ of the fluctuation spectrum should be proportional to the susceptibility of the fluctuating mode  \cite{St71} and is expected to diverge. Nearly in proportion to $P$, it also varied smoothly and monotonically through $V_F$. We are forced to one or the other of two not very palatable conclusions. One of them could be that the observed fluctuations, although influenced by the transition, have an origin that is different from that of the critical Fr\'eedericksz mode. We find this unlikely because the shadowgraph method should reveal an average, orthogonal to the plane, of in-plane director fluctuations that are believed to be associated with the critical mode, and because the time scale of the fluctuations is comparable to the relevant director relaxation-time $\tau_d \simeq 1$ s. Alternatively, the measurements are consistent with a first-order phase transition in the presence of noise sufficiently strong to eliminate  hysteresis and to cause the transition to occur at the thermodynamic transition point where the free energies of the two phases are equal rather than at the point of absolute instability of the ground state.

Our results differ form light scattering measurements on related but different systems that also  have a FT. In a sample with planar alignment and $\epsilon_a > 0$, \cite{WRRZKB91} and in a homeotropic sample in the presence of a magnetic field in the plane, \cite{ELVJ89} $\Gamma_0(V_0)$ was found to vanish at $V_F$. These systems differed from ours in that the ground state below the FT was not rotationally invariant,  either because of the director field \cite{WRRZKB91} or because of the magnetic field \cite{ELVJ89}.  

The rotational invariance of our system below $V_F$ is a feature in common with micro-crystallization of di-block co-polymers \cite{BRFG88} and the onset of Rayleigh-B\'enard convection (RBC) \cite{SH77,OA03}. These two belong to the Brazovskii universality class.\cite{Br75} Members of this class are expected to exhibit a fluctuation-induced first-order transition even though at the mean-field level the transition is of second order. What distinguishes the FT from RBC and the co-polymer case is that it is an instability at $k=0$. In the other systems the rotational invariance is broken at the transition by the selection of a direction in the plane due to the formation of a striped phase with a wave director of finite length; in the present case a spatially uniform domain, corresponding to $k = 0$,  forms above $V_F$ and the symmetry breaking is attributed to the existence of the nematic  director within this domain.

Regardless of the nature of the transition, for macroscopic samples fluctuation effects  should affect the nature of the phase transition only very close to the transition. One can estimate roughly that the relevant thermal noise intensity is given by \cite{RRTSHAB91} $F_{th} \simeq k_B T / k_{33} d$. Here $k_{33} \simeq 8.6\times 10^{-12}$ N is the bend elastic constant.\cite{HDPK92} For our $d = 2.7\times 10^{-5}$ m one has $F_{th} \simeq  2\times 10^{-5}$. The expected critical region has a width of order $\epsilon_c \simeq  F^{2/3} \simeq 7\times 10^{-4}$. \cite{SH77} If we interpret our results as a fluctuation-induced first-order transition, the data suggest $\epsilon_c = {\cal O}(0.1)$ and would require a noise intensity about three orders of magnitude larger than the theoretical estimate. We find it unlikely that there is an experimental noise source that can account for the observations because it would have to be additive (rather than  multiplicative) and it would have to have the appropriate temporal and spatial spectrum.
\begin{figure}
\includegraphics[width=2.4in]{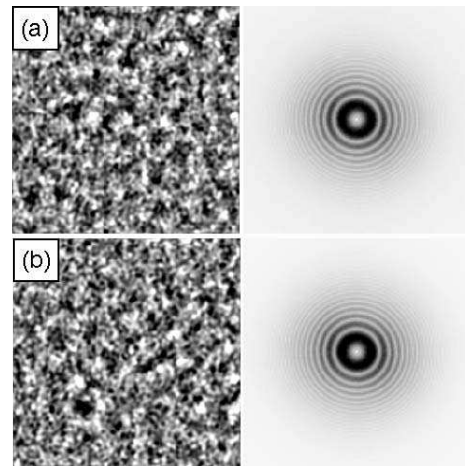}
\caption{Left column: shadowgraph snapshots of an area of $1.03\times1.03$ mm$^2$. Right column: corresponding structure factors, averaged over  256 images. 
(a): $V_0 = 3.170$ , (b): $V_0 = 3.256$ Volt. 
}
\label{fig:images}
\end{figure}

The main sample was N-(4-methoxybenzylidine)-4-butylaniline (MBBA) with $d = 27~\mu$m, doped with less than 0.01\% of tetra-n-butyl-ammonium bromide (TBAB), at 25$^\circ$C driven at $f = 100$ Hz. The same results within our resolution were obtained at $f = 1000$ Hz. The sample was confined between glass plates covered on the inside by ITO electrodes that in turn were coated with a thin film of dimethyloctadecyl[3-(trimethoxysilyl)-propyl]ammonium chloride (DMOAP) to produce homeotropic alignment. The sample had a conductance $\sigma = 5\times 10^{-7}$ (ohm m)$^{-1}$, but similar results were obtained with undoped MBBA and $\sigma = 3\times 10^{-8}$ (ohm m)$^{-1}$. The apparatus, although different in detail, was equivalent to one described before. \cite{DCA98,FN1} Measurements between crossed polarizers for $V_0 > V_F$ showed that the imaged area of $1.03\times 1.03$ mm$^2$ was within a single domain. Time sequences of Images $I_i({\bf x},V_0,\tau),~i = 1,...,N$, $N = 256$ (${\bf x} = (x,y)$ is the two-dimensional position vector in the plane of the sample) were acquired at intervals of three to fifteen seconds with various camera exposure times $0.025 \leq \tau \leq 8$ s at numerous values of $V_0$ over the range $2.5 \alt V_0 \alt 3.6$ Volt. No polarizers were used. At each $V_0$ and $\tau$ each $I_i$ was divided, pixel by pixel, by the average $I_0({\bf x},V_0,\tau)$ of $I_i$. This removed any time-independent structure but left most of the fluctuations. The shadowgraph signal ${\cal I} = I_i/I_0-1$ was Fourier transformed, and the structure factor $S({\bf k},V_0, \tau)$, equal to the square of the modulus of the Fourier transform, was obtained.

Figure~\ref{fig:images} shows grey-scale renderings, rescaled by their own variances, of ${\cal I}$ (left column) and $S$ (right column). Here (a) is well below and (b) is about one percent  above $V_F=3.22$ Volt ($\epsilon \simeq 0.02$). 
There is no obvious change of $\cal I$ and $S$ with $V_0$. It is remarkable  that $S$ is rotationally symmetric even above $V_F$ where the formation of a Fr\'eedericksz domain is known to have broken this symmetry. However, this is similar to the RBC case \cite{OA03} where the fluctuations continue to contribute a {\it ring} in Fourier space even above onset where convection rolls select a direction in the plane.  
\begin{figure}
\includegraphics[width=2.5in]{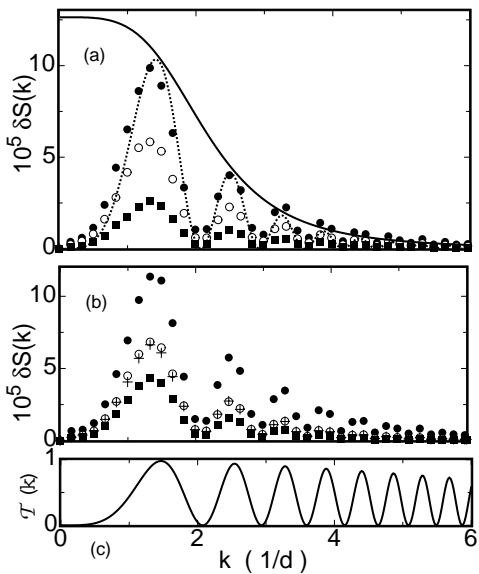}
\caption{(a) and (b): Shadowgraph structure-factor $\delta S(k,V_0)$ as a function of the wave number $k$. (a): camera exposure time $\tau = 0.1$ s and (from top to bottom)  $V_0 = 3.341, 3.256$,  and 3.170 Volt. (b): $V_0 = 3.499$ Volt and $\tau = 0.500$ (solid circles), 4.000 (open circles), and 8.000 s (solid squares). Plusses near open circles: $\tau = 4.000$ s, $f = 1000$ Hz. (c): optical transfer function Eq.~\ref{eq:T_of_k}. Dotted line in (a): fit of Eq.~\ref{eq:S_of_k} to the data for $V_0 = 3.341$ Volt. Solid line: ${\cal B}^2 S_n(k)$ (Eq.~\ref{eq:Sn_of_k}) corresponding to that fit.}
\label{fig:Sofk}
\end{figure}

The radial structure of $S({\bf k},V_0)$ seen in Fig.~\ref{fig:images} is due to the optical transfer function ${\cal T}(k)$ of the shadowgraph. \cite{TC02} In Fig.~\ref{fig:Sofk}a we show the background-corrected azimuthal average $\delta S(k,V_0) = S(k,V_0) - S(k,0)$ of $S({\bf k},V_0)$. Figure~\ref{fig:Sofk}c gives ${\cal T}(k)$ calculated from Eq.~\ref{eq:T_of_k} with the parameters of our optical system. \cite{FN1} It contains the oscillations of $\delta S(k)$ observed in the data. 

We fitted the equations
\begin{eqnarray}
\delta S(k,V_0)&=& {\cal T}(k)S_n(k,V_0)\label{eq:S_of_k}\\
S_n(k,V_0) &=& \frac{B(V_0)}{\xi^4 k^4 + 1}\label{eq:Sn_of_k}\\
{\cal T}(k) = &{\cal B}^2 & \left(\frac{2d}{k z_1 \Theta}\right )^2J_1^2\left (\frac{k z_1 \Theta}{d}\right )sin^2\left (\frac{k^2 z_1}{2 K d^2}\right )\label{eq:T_of_k}
\end{eqnarray}
to the data for $\delta S(k,V_0)$, least-squares adjusting $B$ and $\xi$. Thus, we used the Swift-Hohenberg form Eq.~\ref{eq:Sn_of_k} for the structure factor $S_n$ of the refractive index and the optical transfer function \cite{TC02,BBMTHCA96} in the weak-diffraction limit Eq.~\ref{eq:T_of_k}. In  Eq.~\ref{eq:T_of_k} $z_1$ is the effective optical distance of the shadowgraph, $\Theta=l/2f$ where $l$ is the pinhole size of the light source and $f$ is the focal length of the collimating lens, $K$ is the wave  number of the light,  ${\cal B} = 2 d K$, and $J_1$ is the first-order Bessel function of the first kind.\cite{BBMTHCA96,FN1} 
As an example we show the fit to the data for $V_0 = 3.341$ Volt in Fig.~\ref{fig:Sofk}a as a dotted line. The solid line in Fig.~\ref{fig:Sofk}a gives the corresponding result for ${\cal B}^2S_n$ from Eq.~\ref{eq:Sn_of_k}.

\begin{figure}
\includegraphics[width=2in]{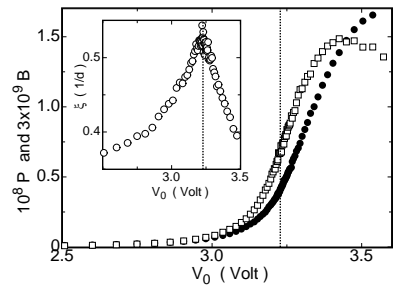}
\caption{The amplitude $B$ (open squares) and power $P = \pi^2 B / 2 \xi^2$ (solid circles) of the refractive-index structure factor $S_n$, as a function of the voltage amplitude $V_0$. The insert gives the correlation length $\xi$ [inverse width at half height of $S_n(k)$] as a function of $V_0$. Dotted lines: $V_F = 3.22$ Volt.}
\label{fig:power}
\end{figure}

Results for $B$ are shown in Fig.~\ref{fig:power} as open squares. They vary smoothly through the transition, continue to rise beyond it, and finally have a maximum about 10\% above $V_F$. A very different picture emerges from the data for $\xi$ that are shown in the insert in Fig.~\ref{fig:power}. We see that $\xi$ has a sharp cusp at $V_F = 3.22$ Volt, indicating that the transition is sharp roughly at the one percent level. However, there is no divergence of $\xi$ as would be expected for a second-order phase transiton. From $\xi$, $B$, and Eq.~\ref{eq:Sn_of_k} we can compute the total refractive-index fluctuation-power $P = 2\pi \int_0^\infty k S_n(k) dk = \pi^2 B / 2 \xi^2$. This is shown as  solid circles in Fig.~\ref{fig:power}. The $V_0$-dependence of $P$ is similar to that of $B$; no singularity at $V_F$ is apparent from the data. Beyond $V_0 \simeq 3.6$ $P$ decreases again.
\begin{figure}
\includegraphics[width=2in]{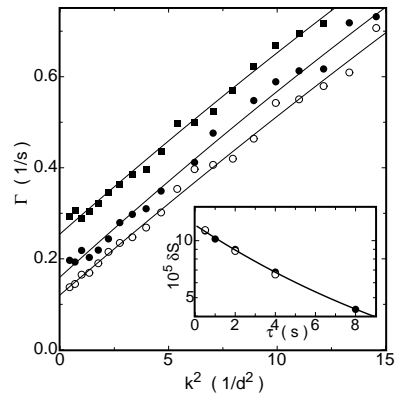}
\caption{The relaxation rate $\Gamma(k)$ as a function of the square of the wave number $k$. Solid circles: $V_0 = 3.50$ Volt. Open circles: $V_0 = 3.25$ Volt. Solid squares: $V_0 = 3.00$ Volt. The lines are fits of $\Gamma = \Gamma_0 + \Gamma_2 k^2 + \Gamma_4k^4$ to the data. Insert: example of $\delta S(k,V_0,\tau)$ on a logarithmic scale as a function of $\tau$ on a linear scale for $V_0 = 3.50$ Volt, $k = 1.327$, and $f = 100$ Hz (solid circles) and 1000 Hz (open circles). Solid line: fit of Eq.~\ref{eq:Softau} to the 100 Hz data.
}
\label{fig:Gamma}
\end{figure}
\begin{figure}
\includegraphics[width=2.5in]{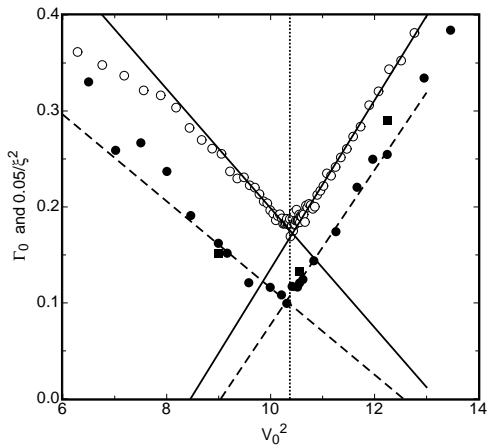}
\caption{The relaxation rate $\Gamma_0$ (solid circles: f = 100 Hz; solid squares: f = 1000 Hz) and the inverse square of the correlation length $1/\xi^2$ (open circles, f=100 Hz) as a fuction of $V_0^2$.  Both have a cusp but remain non-zero  at $V_F$ The lines are straight-line fit to the data near the transition.}
\label{fig:Gamma0}
\end{figure}

In order to learn about the dynamics of the fluctuations, we took images with several camera exposure times $\tau$ at each of several values of $V_0$. Larger values of $\tau$ lead to more averaging of the fluctuations and thus to a smaller $S(k,V_0,\tau)$. This is illustrated by the results shown in Fig.~\ref{fig:Sofk}b. We are not aware of a prediction of the form of the time-correlation function $C(k, \delta t)$ of the fluctuations in a thin nematic sample and for $V_0 > 0$, such as was done for RBC \cite{OOSA04}. Thus we assume that $C(k,\delta t)$ is well approximated by a simple exponential decay $exp[-\Gamma(k) \delta t]$. One can show that this leads to  
\begin{equation}
\delta S(k,V_0,\tau) = \delta S(k,V_0,0)\frac{1-exp[-\Gamma(k,V_0)\tau]}{\Gamma(k,V_0)\tau}
\label{eq:Softau}
\end{equation}
for the average over a time interval $\tau$. At each of several $V_0$, we fitted Eq.~\ref{eq:Softau} to measurements of  $\delta S(k,V_0,\tau)$ at each of many values of $k$ to give $\delta S(k,V_0,0)$ and $\Gamma(k,V_0)$. An example is shown in the insert of Fig.~\ref{fig:Gamma}. Although $\delta S(k,V_0,0)$ is influenced by ${\cal T}(k)$, we note that $\Gamma(k,V_0)$ is not. 
In Fig.~\ref{fig:Gamma} we show some of the results for $\Gamma$ as a function of $k^2$. The data can be represented well by $\Gamma = \Gamma_0 + \Gamma_2 k^2 + \Gamma_4k^4$ (solid lines in the figure). 
Results for  $\Gamma_0$ are given in Fig.~\ref{fig:Gamma0}. One sees that $\Gamma_0$ has a sharp down-pointing cusp at $V_F$ (vertical dotted line); but contrary to predictions for a second-order phase transition $\Gamma_0$ remains finite. 
Figure~\ref{fig:Gamma0} also shows the results for $\xi$ in the form of $0.05/\xi^2$ as a function of $V_0^2$. For a second-order transition and near $V_F$ one would expect $\xi^{-2}$ to be a linear function of $V_0^2$ and to vanish at $V_F$. However, $\xi^{-2}$ behaves very similarly to $\Gamma_0$ and remains finite. The solid and dashed lines  are straight-line fits to the data near $V_F^2$, separately above and below the transition. The fits for $V_0^2 < V_F^2$ extrapolate approximately to $\Gamma_0 = 0$ and $1/\xi^2 = 0$ at $V_c^2 \simeq 13~(V_c \simeq 3.6)$, consistent with a point of absolute instability at $V_c$ that, for a first-order transition, can not be reached in the presence of strong noise. Interestingly, $B$ and $P$ (Fig.~\ref{fig:power}) reach their maxima near $V_c$.

We presented data for both the statics and the dynamics of the bend Fr\'eedericksz transition of a homeotropically aligned NLC that are inconsistent with a second-order phase transition and suggestive of a first-order transition in the presence of strong noise. This disagrees with our theoretical understanding of this system that predicts a second-order transition  at the mean-field level. \cite{HDPK92} A fluctuation-induced first-order transition \cite{Br75} should manifest itself only in a very small region, \cite{SH77} within 0.1 percent or so of the mean-field transition, whereas the data suggest a critical region of width  $\epsilon_c = {\cal O}(0.1)$.   

This work was supported by the National Science Foundation through Grant DMR02-43336.


\begin{references}
\bibitem{FN1}We used a SUNLED LZE12W light-emitting diode with $K = 9.97\times 10^4$ cm$^{-1}$. Other parameters are $d = 27 \mu$m, $z_1 = 1.06$ cm, $l = 50 \mu$m, $f = 5$ cm, and thus $\Theta = 5\times 10^{-4}$ and ${\cal B} = 538$.
\end{references}
\end{document}